\documentclass[prd,showpacs,amsmath,amssymb,twocolumn]{revtex4}
\usepackage{graphicx,color,slashbox}
\begin{document}


\title{What can we learn from the decay of $ N_X(1625)$ in molecule picture?}

\vspace*{1.0cm}

\author{Xiang
Liu}\email{xiangliu@pku.edu.cn}\author{Bo
Zhang}\affiliation{Department of physics, Peking University,
Beijing, 100871, China}

\vspace*{1.0cm}

\date{\today}

\begin{abstract}
Considering two molecular state assumptions, i.e. S-wave
$\bar{\Lambda}-K^-$ and S-wave $\bar{\Sigma}^0-K^-$ molecular
states, we study the possible decays of $\bar N_X(1625)$ that
include $\bar N_X(1625)\to K^{-}\bar{\Lambda}, \pi^{0}\bar{p},
\eta\bar{p}, \pi^{-}\bar{n}$. Our results indicate: (1) if $\bar
N_{X}(1625)$ is $\bar{\Lambda}-K^-$ molecular state,
$K^{-}\bar{\Lambda}$ is the main decay modes of $\bar N_{X}(1625)$,
and the branching ratios of the rest decay modes are tiny; (2) if
$\bar N_{X}(1625)$ is $\bar{\Sigma}^0-K^-$ molecular state, the
branching ratio of $\bar N_{X}(1625)\to K^{-}\bar{\Lambda}$ is one
or two order smaller than that of $\bar N_{X}(1625)\to
\pi^{0}\bar{p}, \eta\bar{p}, \pi^{-}\bar{n}$. Thus the search for
$\bar N_X(1625)\to \pi^{0}\bar{p}, \eta\bar{p}, \pi^{-}\bar{n}$ will
be helpful to shed light on the nature of $\bar N_X(1625)$.
\end{abstract}

 \maketitle

\section{Introduction}\label{sec1}

Two years ago, BES Collaboration announced an enhancement
$\bar{N}_X(1625)$ by studying the $K^{-}\bar{\Lambda}$ invariant
mass spectrum in $J/\psi\to pK^{-}\bar{\Lambda}$ channel
\cite{HXYANG,JIN,SHEN}. BES Collaboration gave the rough measurement
result about the mass and width of $\bar{N}_X(1625)$:
$m=1500\sim1650$ MeV, $\Gamma=70\sim110$ MeV. Experiment also
indicates that spin-parity favors $\frac{1}{2}^-$ for ${N}_X(1625)$,
which denotes the antiparticle of $\bar{N}_X(1625)$.

Using the branching ratio $B(J/\psi\to p\bar{p})=2.17\times 10^{-3}$
\cite{PDG} as a reference, we can deduce $B[\bar N_{X}(1625)\to
\bar{\Lambda}K^{-}]\sim 10\%$ if $\bar N_{X}(1625)$ is a regular
baryon and the branching ratio of $J/\psi\to p\bar{N}_{X}(1625)$
should be comparable with that of $J/\psi\to p\bar{p}$, which shows
that there exists strong coupling between $\bar{N}_{X}(1625)$ and
$K^{-}\bar{\Lambda}$.

This enhancement structure inspired several theoretical speculations
of its underlying structure. The authors of Ref. \cite{ZHANG}
studied the S-wave $\Lambda K$ and $\Sigma K$ with isospin $I= 1/2$
within the framework of the chiral $SU(3)$ quark model by solving a
resonating group method (RGM) equation. Their results show a strong
attraction between the $\Sigma$ and $K$, and a $\Sigma K$ quasibound
state is thus formed as a consequence with a binding energy of about
17 MeV, whereas the $\Lambda K$ is unbound. Considering small mass
difference of the $\Lambda K$ and $\Sigma K$ thresholds, the strong
attraction between $\Sigma$ and $K$, and the sizable off-diagonal
matrix elements of $\Lambda K$ and $\Sigma K$, they also
investigated the coupled channel effect of $\Lambda K$ and $\Sigma
K$, and found that a sharp resonance with a mass $M=1669$ MeV and a
width $\Gamma=5$ MeV \cite{ZHANG}.

Liu and Zou suggested that enhancement structure $\bar N_X(1625)$
comes from the strong coupling between $\bar N(1535)$ and
$K\Lambda$. Furthermore $R=g_{\bar N(1535)K\Lambda}/g_{\bar
N(1535)p\eta}$ are extracted by the branching ratios taken from
BES experiments on $J/\psi\to \bar{p}p\eta$
\cite{peta-1,peta-2,peta-3} and $J/\psi\to
 {p}K^- \bar{\Lambda}$ \cite{HXYANG}. The new obtained value
of $g_{\bar{N}(1535)K^-\bar{\Lambda}}$ is shown to reproduce recent
$pp\to {p}K^- \bar{\Lambda}$ near-threshold cross section data as
well \cite{LIU}.

At recent Hadron 07 conference, BES Collaboration reported the
preliminary new experiment result of $\bar N_{X}(1625)$. Its mass
and width are well determined as \cite{BES-N(1625)}
$$m=1625^{+5+13}_{-7-23}\;\;\mathrm{MeV},\;\; \Gamma=43^{+10+28}_{-7\;\;-11}\;\;\mathrm{MeV}$$ respectively.
The production rate of $\bar{N}_X(1625)$ is
\begin{eqnarray*}
&&B[J/\psi\to p \bar{N}_{X}(1625)]\cdot B[\bar{N}_{X}(1625)\to
K^{-}\bar{\Lambda}]\\&&=9.14^{+1.30+4.24}_{-1.25 -8.28}\times
10^{-5}.
\end{eqnarray*}
These more accurate experimental information of $\bar N_X(1625)$
provides us good chance to further study the nature of $\bar
N_X(1625)$.

Despite two theoretical speculations proposed above, at present the
study of decays of $\bar N_X(1625)$, which play an important role to
clarify the properties of $\bar N_X(1625)$, is missing. In this
work, we firstly assume $\bar N_{X}(1625)$ to be a molecular state
and is dedicated to the study of the possible decays of $\bar
N_X(1625)$. For the convenience of comparing with BES experiment,
one focuses on the study of decays of antiparticle $\bar{N}_X(1625)$
with the spin-parity $\frac{1}{2}^{+}$.

This paper is organized as follow. In Sect. \ref{sec2}, we present
the formulation about the possible decays of $\bar{N}_X(1625)$. In
Sect. \ref{sec3}, the numerical results are given. The last
section is the conclusion and discussion.

\section{FORMULATION}\label{sec2}



In this work we do not focus on whether $\bar{\Lambda}-K^-$ or
$\bar{\Sigma}^{0}-K^-$ can form S-wave molecular state, which is
investigated in Ref. \cite{ZHANG}. Whereas we are mainly dedicated
to the study of possible decays of $\bar{N}_X(1625)$ in two
different assumptions of molecular states.


\subsection{The possible decays assuming $\bar{N}_X(1625)$ to be
$\bar{\Lambda}-K^-$ molecular state}

\begin{figure}[htb]
\begin{center}
\scalebox{0.5}{\includegraphics{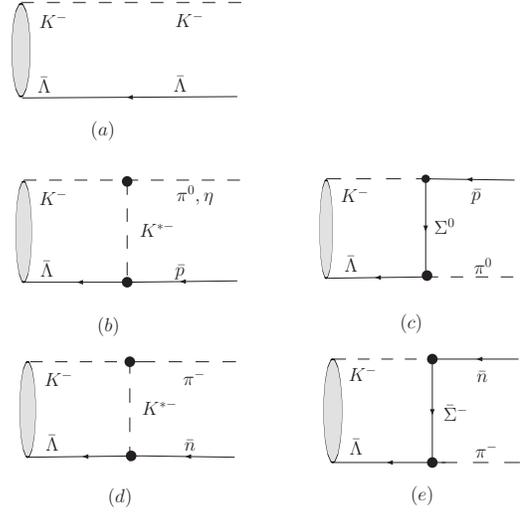}}
\end{center} \caption{The diagrams depicting the decays of $\bar{N}_X(1625)$
in the picture of $\bar{\Lambda}-K^-$ molecular
state.}\label{hadron-1}
\end{figure}

In the assumption of $\bar{\Lambda}-K^-$ molecular state, the most
direct decay mode of $\bar{N}_X(1625)$ is $\bar{N}_X(1625)\to
\bar{\Lambda}+K^-$ depicted in Fig. \ref{hadron-1} (a). Its decay
amplitude is
\begin{eqnarray}
\mathcal{M}[\bar{N}_X(1625)\to
\bar{\Lambda}+K^-]=i\mathcal{G}\bar{v}_{N}\gamma_{5}{v}_{\bar{\Lambda}},
\end{eqnarray}
where $\mathcal{G}$ is the coupling constant between
$\bar{N}_X(1625)$ and $\bar{\Lambda}K^-$. $v_{\bar{\Lambda}}$ and
$v_{N}$ are the spinors.

Besides the direct decay, there are several subordinate decays
depicted in Fig. \ref{hadron-1} (c)-(e) by the final state
interaction (FSI) effect. For obtaining their decay amplitudes,
one needs to use the below Lagrangians \cite{swart,lagrangian}:
\begin{eqnarray}
\mathcal{L}_{\mathcal{PPV}}&=&-ig_{_{\mathcal{PPV}}}Tr\big([\mathcal{P},\partial_{\mu}\mathcal{P}]\mathcal{V}^{\mu}\big),\label{ppv}\\
\mathcal{L}_{\mathcal{BBP}}&=&F_{P}Tr\big(\mathcal{P}[\mathcal{B},\bar{\mathcal{B}}]\big)\gamma_{5}+D_{P}Tr\big(\mathcal{P}\{\mathcal{B},\bar{\mathcal{B}}\}
\big)\gamma_{5},\label{BBP}\\
\mathcal{L}_{\mathcal{BBV}}&=&
F_{V}Tr\big(\mathcal{V}^{\mu}[\mathcal{B},\bar{\mathcal{B}}]\big)\gamma_{\mu}
+D_{V}Tr\big(\mathcal{V}^{\mu}\{\mathcal{B},\bar{\mathcal{B}}\}\big)\gamma_{\mu},\nonumber\label{BBV}\\
\end{eqnarray}
where the concrete values of the coupling constants will be given
in detail in the following section. $\bar{B}$ is the Hermitian
conjugate of $B$. $\mathcal{P}$, $\mathcal{V}$ and $B$
respectively denote the octet pseudoscalar meson, the nonet vector
meson and baryon matrices:
\begin{eqnarray*}
\mathcal{P}&=&\left(\begin{array}{ccc}
\frac{\pi^{0}}{\sqrt{2}}+\frac{{\eta}}{\sqrt{6}}&\pi^{+}&K^{+}\\
\pi^{-}&-\frac{\pi^{0}}{\sqrt{2}}+\frac{\eta}{\sqrt{6}}&
K^{0}\\
K^- &\bar{K}^{0}&-\frac{2}{3}\eta
\end{array}\right),\\
\mathcal{V}&=&\left(\begin{array}{ccc}
\frac{\rho^{0}}{\sqrt{2}}+\frac{\omega}{\sqrt{2}}&\rho^{+}&K^{*+}\\
\rho^{-}&-\frac{\rho^{0}}{\sqrt{2}}+\frac{\omega}{\sqrt{2}}&
K^{*0}\\
K^{*-} &\bar{K}^{*0}&\phi
\end{array}\right),\\
\mathcal{B}&=&\left(\begin{array}{ccc}
\frac{\Sigma^{0}}{\sqrt{2}}+\frac{\Lambda}{\sqrt{6}}&\Sigma^{+}&p\\
\Sigma^{-}&-\frac{\Sigma^{0}}{\sqrt{2}}+\frac{\Lambda}{\sqrt{6}}&
n\\
\Xi^{*-} &\Xi^{*0}&\frac{-2\Lambda}{\sqrt{6}}
\end{array}\right).
\end{eqnarray*}

Because $M_{\bar{\Lambda}}+M_{K^-}$ is about 1610 MeV, which is
less than the mass of $\bar{N}_{X}(1625)$, thus intermediate
states $\bar{\Lambda}$ and $K^{-}$ in Fig. \ref{hadron-1} (b)-(d)
can be on-shell. By Cutkosky cutting rules, one writes out the
general amplitude expression corresponding to Fig. \ref{hadron-1}
(b), (d)
\begin{eqnarray}
\mathcal{M}_{1}^{(\mathcal{A}_1,\mathcal{C}_1)}&=&\frac{1}{2}\int\frac{d^3
p_{1}}{(2\pi)^{3}2E_{1}}\frac{d^3
p_{2}}{(2\pi)^{3}2E_{2}}\nonumber\\
&&\times(2\pi)^{4}\delta^{4}(M_{N}-p_{1}-p_{2})[i\mathcal{G}\bar{v}_{N}\gamma_{5}{v}_{\bar{\Lambda}}]
\nonumber\\
&&\times
[ig_{1}\bar{v}_{{\bar{\Lambda}}}\gamma_{\mu}v_{{\mathcal{A}_1}}][ig_{2}(p_{1}+p_{3})_{\nu}]\frac{i}{q^{2}-M_{\mathcal{C}_1}^{2}}
\nonumber\\
&&\times\Big[-g^{\mu\nu}+\frac{q^{\mu}q^{\nu}}{M_{\mathcal{C}_1}^2}\Big]\mathcal{F}^{2}(M_{\mathcal{C}_1},q^2).
\end{eqnarray}
For Fig. \ref{hadron-1} (c), (e), the general amplitude expression
is
\begin{eqnarray}
\mathcal{M}_{1}^{(\mathcal{A}_2,\mathcal{C}_2)}&=&\frac{1}{2}\int\frac{d^3
p_{1}}{(2\pi)^{3}2E_{1}}\frac{d^3
p_{2}}{(2\pi)^{3}2E_{2}}\nonumber\\
&&\times(2\pi)^{4}\delta^{4}(M_{N}-p_{1}-p_{2})[i\mathcal{G}\bar{v}_{N}\gamma_{5}v_{\bar{\Lambda}}]
\nonumber\\
&&\times [ig_{2}'\bar{v}_{\bar{\Lambda}}\gamma_{5}]
\frac{i(q\!\!\!\slash+M_{\mathcal{C}_2})}{q^{2}-M_{\mathcal{C}_2}^{2}}[ig_{1}'\gamma_{5}{v}_{\mathcal{A}_2}]
\nonumber\\
&&\times\mathcal{F}^{2}(M_{\mathcal{C}_2},q^2).
\end{eqnarray}
In the above expressions, $\mathcal{C}_{i}$ and $\mathcal{A}_{i}$
denote the exchanged particle and the final state baryon
respectively. $p_{1}$ and $p_{2}$ are respectively the four
momenta of $K^-$ and $\bar{\Lambda}$. $\mathcal{F}^{2}(m_{i},q^2)$
etc denotes the form factor which compensates the off-shell
effects of hadrons at the vertices. In this work, one takes
$\mathcal{F}^{2}(m_{i},q^2)$ as the monopole form
\cite{FF,HY-Chen}
\begin{eqnarray}
\mathcal{F}^{2}(m_{i},q^2)=\bigg(\frac{\xi^{2}-m_{i}^2
}{\xi^{2}-q^{2}}\bigg)^2,\label{form}
\end{eqnarray}
where $\xi$ is a phenomenological parameter. As $q^2\to 0$ the form
factor becomes a number. If $\xi\gg m_{i}$, it becomes unity. As
$q^2\rightarrow\infty$, the form factor approaches to zero. As the
distance becomes very small, the inner structure would manifest
itself and the whole picture of hadron interaction is no longer
valid. Hence the form factor vanishes and plays a role to cut off
the end effect. The expression of $\xi$ is \cite{HY-Chen}
\begin{eqnarray}
\xi(m_{i})=m_{i}+\alpha \Lambda_{QCD},
\end{eqnarray}
where $m_{i}$ denotes the mass of exchanged meson.
$\Lambda_{QCD}=220$ MeV. $\alpha$ is a phenomenological parameter
and is of order unity.

\subsection{The decay modes assuming $\bar{N}_{X}(1625)$ to be a
$\bar{\Sigma}^0-K^-$ molecular state}

Because of having no enough phase space, $\bar{N}_{X}(1625)$ can not
decay to $\bar{\Sigma}^0$ and $K^-$.

\begin{figure}[htb]
\begin{center}
\scalebox{0.5}{\includegraphics{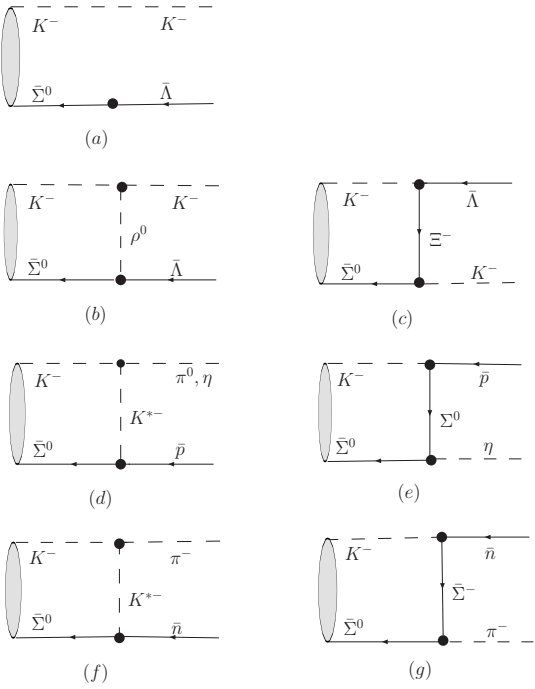}}
\end{center} \caption{The diagrams depicting the decays
of $\bar{N}_{X}(1625)$ in the assumption of $\bar{\Sigma}^0-K^-$
molecular state for $\bar{N}_{X}(1625)$.}\label{hadron-2}
\end{figure}

Isospin violation effect can result in the mixing of $\Sigma$ with
$\Lambda^{0}$ \cite{s.l. zhu}. Thus decay $\bar{N}_X(1625)\to
\bar{\Lambda}+K^-$ occurs, which is depicted by Fig. \ref{hadron-2}
(a). Using the Lagrangian
$$\mathcal{L}_{\mbox{mixing}}=g_{\mbox{mixing}}(\bar{\psi}_{{\Sigma}^0}\psi_{{\Lambda}}+\bar{\psi}_{{\Lambda}}{\psi}_{{\Sigma}^0})$$
with the coupling constant $\theta=0.5\pm 0.1$ MeV obtained by QCD
sum rule \cite{s.l. zhu}, one obtains the decay amplitude
\begin{eqnarray}
\mathcal{M}[\bar{N}_X(1625)\to
\bar{\Sigma}^{0}+K^-]=\mathcal{G}\;g_{\mbox{mixing}}\;
\bar{v}_{N}\gamma_{5}\frac{i}{p\!\!\!\slash-M_{\Lambda}}v_{\bar{\Lambda}},\nonumber\\
\end{eqnarray}
where $p$ and $M_{\Lambda}$ are the four momentum and mass carried
by $\bar{\Lambda}$.

For the decays depicted in Fig. \ref{hadron-2} (b)-(g),
$\bar{\Sigma}^0$ and $K^-$ are off-shell. Thus the general
expression of Fig. \ref{hadron-2} (b), (d), (f) is expressed as
\begin{eqnarray}
\mathcal{M}_{3}^{(\mathcal{A}_3,\mathcal{C}_3)}&=&\int\frac{d^4
q}{(2\pi)^{4}}[i\mathcal{G}\bar{v}_{N}\gamma_{5}]
\frac{i}{-\not{p}_{2}-M_{\bar{\Sigma}^{0}}}[ig_{3}\gamma_{\mu}{v}_{{\mathcal{A}_3}}]
 \nonumber\\&&
\times[ig_{4}(p_{1}+p_{3})_{\nu}]\frac{-ig^{\mu\nu}}{q^{2}-M_{\mathcal{C}_3}^{2}}
\frac{i}{p_{1}^{2}-M_{K}^2}\nonumber\\&&
\times\mathcal{F}^{2}(M_{\mathcal{C}_3},q^2),\label{hadron-loop-1}
\end{eqnarray}
for Fig. \ref{hadron-2} (c), (e), (g) the general amplitude
expression reads as
\begin{eqnarray}
\mathcal{M}_{4}^{(\mathcal{A}_4,\mathcal{C}_4)}&=&\int\frac{d^4
q}{(2\pi)^{4}}[i\mathcal{G}\bar{v}_{N}\gamma_{5}]\frac{i(\not{p}_{2}-M_{\bar{\Sigma}^0})}{-p_{2}^{2}-M_{\bar{\Sigma}^0}^{2}}[ig_{4}'\gamma_{5}]
\nonumber\\
&&\times
\frac{i(q\!\!\!\slash+M_{\mathcal{C}_4})}{q^{2}-M_{\mathcal{C}_4}^{2}} [ig_{3}'\gamma_{5}{v}_{\mathcal{A}_4}]\nonumber\\
&&\times\frac{i}{p_{1}^{2}-M_{K}^{2}}
\mathcal{F}^{2}(M_{\mathcal{C}_4},q^2),\label{hadron-loop-2}
\end{eqnarray}
where $p_{1}$ and $p_2$ denote the four momenta carried by $K^-$
and $\bar{\Sigma}^0$ respectively. $q=p_1-p_3=p_4-p_2$. The
definition of $\mathcal{F}^{2}(m_i,q^2)$ is given in eq.
(\ref{form}). Moreover the form factor may provide a convergent
behavior for the triangle loop integration. That is very similar
to the case of the Pauli-Villas renormalization scheme
\cite{Izukson,peskin}.

Using the same treatment in Ref. \cite{liuxiang}, we obtain the
further expressions of eqs. (\ref{hadron-loop-1}) and
(\ref{hadron-loop-2}) that are listed in appendix.

\section{numerical results}\label{sec3}

In QCD sum rule approach, the ratios of coupling constants in eqs.
(\ref{BBP}) and (\ref{BBV}) are given as $F_{P}/D_{P}=0.6$
\cite{F/D} and ratio $F_{V}/(F_{V}+D_{V})=1$ \cite{thesis}. In the
limit of SU(3) symmetry, by $g_{NN\pi}=13.5$ and $g_{NN\rho}=3.25$
\cite{coupling}, one obtains the meson-baryon coupling constants
relevant to our calculation: $g_{PPV}=6.1$, $F_P=13.5$, $D_P=0$,
$F_V=1.2$, $D_V=2.0$.\\

\begin{figure}[htb]
\begin{center}
\scalebox{0.8}{\includegraphics{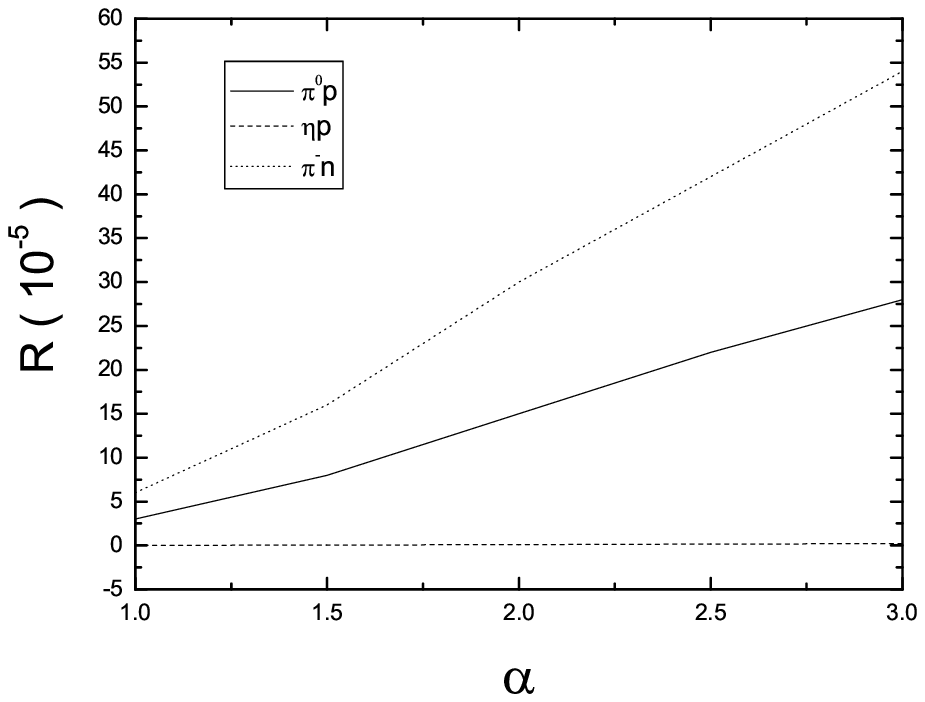}}
\end{center} \caption{The ratios of $\bar N_X(1625)\to\pi^0\bar{p},\eta\bar{p},\pi^{-}\bar{n}$
decay widths to $\bar{N}_X(1625)\to \bar{\Lambda} K^-$ decay width
in the picture of $\bar{\Lambda}-K^-$ molecular
state.}\label{ratio-1}
\end{figure}

\begin{figure}[htb]
\begin{center}
\scalebox{0.8}{\includegraphics{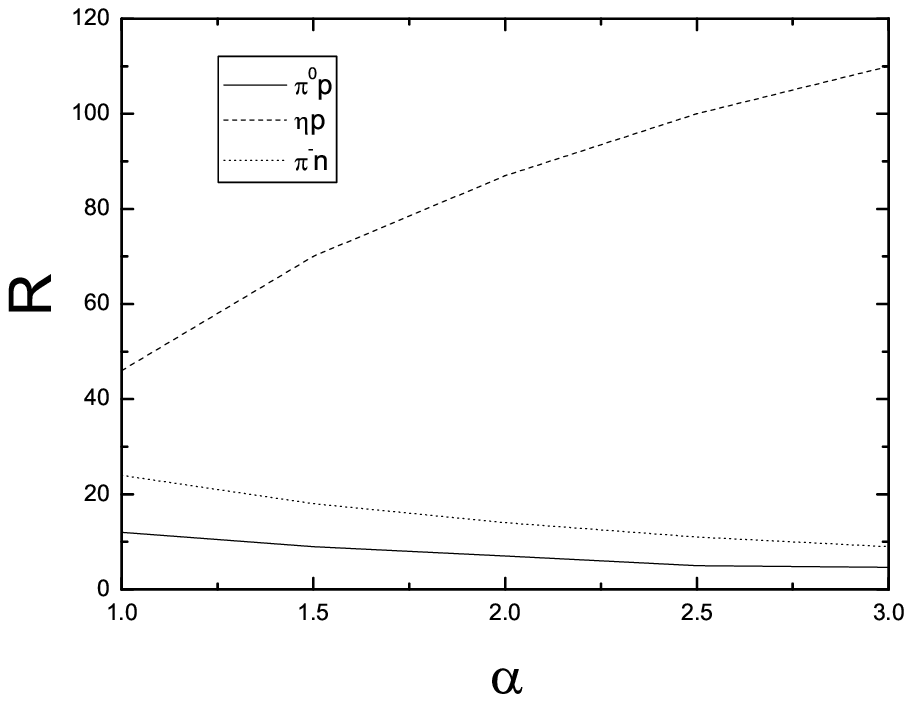}}
\end{center} \caption{The ratios of $\bar N_X(1625)\to\pi^0\bar{p},\eta\bar{p},\pi^{-}\bar{n}$
decay widths to $\bar N_X(1625)\to \bar{\Lambda} K^-$ decay width in
$\bar{\Sigma}^0-K^-$ molecular state picture.}\label{ratio-2}
\end{figure}

Using the above parameters as input, we get the ratios of the decay
widths of
$\bar{N}_X(1625)\to\pi^0\bar{p},\eta\bar{p},\pi^{-}\bar{n}$ to the
decay width of $\bar N_X(1625)\to \bar{\Lambda} K^-$ in the
assumptions of $\bar{\Lambda}-K^-$ molecular state and
$\bar{\Sigma}^0-K^-$ molecular state for $\bar{N}_X(1625)$, which
are shown in Fig. \ref{ratio-1} and Fig. \ref{ratio-2} respectively.
Here $\alpha$ in the form factor is taken as the range $1\sim 3$
\cite{HY-Chen}.

Fig. \ref{ratio-1} and Fig. \ref{ratio-2} show that these ratios
do not strongly depend on the $\alpha$. By taking a typical value
$\alpha=1.5$, one further gives the following ratios listed in
Table \ref{ratio-all}.
\begin{center}
\begin{table}[htb]
\begin{tabular}{c|ccc}\hline
&$\frac{\Gamma(\pi^{0}\bar{p})}{\Gamma(K^{-}\bar{\Lambda})}$&$\frac{\Gamma(\eta\bar{p})}{\Gamma(K^{-}\bar{\Lambda})}
$&$\frac{\Gamma(\pi^{-}\bar{n})}{\Gamma(K^{-}\bar{\Lambda})}$\\\hline
$\bar{\Lambda}-K^-$&$1\times 10^{-4}$&$5\times10^{-7}$&$2\times
10^{-4}$\\\hline $\bar{\Sigma}^0-K^-$&9&$70$&18\\\hline
\end{tabular}\caption{The ratios of the
decay widths of
$\bar{N}_X(1625)\to\pi^0\bar{p},\eta\bar{p},\pi^{-}\bar{n}$ to the
decay width of $\bar{N}_X(1625)\to \bar{\Lambda} K^-$ in different
molecular assumptions with $\alpha=1.5$.\label{ratio-all}}
\end{table}
\end{center}

By using these ratios shown in Figs. \ref{ratio-1}, \ref{ratio-2}
and the branching ratio $B[J/\psi\to p \bar{N}_{X}(1625)]\cdot
B[\bar{N}_{X}(1625)\to K^{-}\bar{\Lambda}]=9.14^{+1.30+4.24}_{-1.25
-8.28}\times 10^{-5}$ given by BES \cite{BES-N(1625)}, one estimates
the branching ratio of subordinate decays of $\bar{N}_X(1625)$ in
$J/\psi$ decay shown in Table. \ref{numerical}. Due to the
uncertainty of $\alpha$, thus we given the possible ranges for these
branching ratios.

If $\bar{N}_{X}(1625)$ is $\bar{\Lambda}-K^-$ molecular state,
$\bar{N}_X(1625)$ mainly decay to $K^{-}\bar{\Lambda}$. The
branching ratios of the subordinate decays
$\bar{N}_X(1625)\to\pi^{0}\bar{p},\eta\bar{p},\pi^{-}\bar{n}$ are
far less than that of $\bar N_X(1625)\to K^{-}\bar{\Lambda}$, which
can explain why $\bar N_X(1625)$ was firstly observed in the mass
spectrum of $K^{-}\bar{\Lambda}$. In the Particle Data Book
\cite{PDG}, the smallest branching ratios that have been measured
for $J/\psi$ decays are larger than $10^{-5}$. Thus the rest decays
of $\bar N_X(1625)$ is hardly measured in further experiments.

If $\bar N_{X}(1625)$ is S-wave $\bar{\Sigma}^{0}-K^-$ molecular
state, $\bar N_X(1625)$ can not decay to $\bar{\Sigma}^{0}K^-$ due
to having no enough phase space. Because of the $\Lambda-\Sigma^0$
mixing mechanism and final state interaction effect, $\bar
N_X(1625)$ can decay to $\bar{\Lambda}K^-$. Our calculations
indicate that the branching ratio of $\bar
N_X(1625)\to\bar{\Lambda}K^-$ is about one or two order smaller than
that of $\bar N_X(1625)\to \pi^{0}\bar{p}, \eta\bar{p},
\pi^{-}\bar{n}$. Although the neutral particles in the decay modes
$\pi^{0}\bar{p}, \eta\bar{p}, \pi^{-}\bar{n}$ increase the
difficulty of searching these decay modes in experiment, future
experiments still have the potential to find $\bar N_X(1625)\to
\pi^{0}\bar{p}, \eta\bar{p}, \pi^{-}\bar{n}$.

\section{Discussion and conclusion}\label{sec4}

In this work, we focus on different results of the decay mode of
$\bar{N}_{X}(1625)$ resulted from two molecule assumptions, i.e.
S-wave $\bar{\Lambda}-K^-$ and S-wave $\bar{\Sigma}^{0}-K^-$
systems. Basing on these two pictures, we estimate the possible
decay modes of $\bar{N}_X(1625)$, which include
$K^{-}\bar{\Lambda}$, $\pi^{0}\bar{p}$, $\eta\bar{p}$ and
$\pi^{-}\bar{n}$. Our result indicates that the search for $\bar
N_X(1625)\to \pi^{0}\bar{p}, \eta\bar{p}, \pi^{-}\bar{n}$ will
shed light on the nature of $\bar N_X(1625)$.

At present the experimental information indicates that $\bar
N_X(1625)$ is of very strong coupling with $\bar{\Lambda}K^{-}$,
and other modes is still missing \cite{BES-N(1625)}. Thus the
assumption of S-wave $\bar{\Lambda}-K^-$ molecular state is more
favorable than that of S-wave $\bar{\Sigma}^0-K^-$ molecular state
for $\bar N_X(1625)$. However, the result of Ref. \cite{ZHANG}
indicates that it is difficult to form a $\Lambda K$ bound state.
In fact, in molecule picture, in general such an S-wave
$\bar{\Lambda}K^{-}$ system should be of very wide width, which
contradicts with the experimental information of
$\bar{N}_{X}(1625)$ ($\Gamma_{\bar{N}_{X}(1625)}=43$ MeV).
Although the above analysis shows that S-wave $\bar{\Lambda}-K^-$
molecule assignment as $\bar N_X(1625)$ is not suitable, we still
try to study the decay of $\bar{N}_X(1625)$ in S-wave
$\bar{\Lambda}-K^-$ molecule picture.

In the assumption of S-wave $\bar{\Sigma}^0-K^-$ molecular state,
the sum of branching ratios of $\bar N_X(1625)\to \pi^{0}\bar{p},
\eta\bar{p}, \pi^{-}\bar{n}$ listed in Table \ref{numerical} is
about $10^{-2}$. Such large branching ratio is unreasonable for
$J/\psi$ decay. BES collaboration has already studied $J/\psi\to
p\pi^-\bar{n}$ in Ref. \cite{peta-2} and $J/\psi\to
p(\eta\bar{p})$ in Ref. \cite{peta-3}. The branching ratios
respectively corresponding to $J/\psi\to p\pi^-\bar{n}$ and
$J/\psi\to p\eta\bar{p}$ are $2.4\times 10^{-3}$ and $2.1\times
10^{-3}$ \cite{peta-2,peta-3}. Although these experimental values
is comparable with our numerical result of corresponding channel,
experiments did not find structure consistent with
$\bar{N_{X}}(1625)$, which seem to show that evidence against
S-wave $\bar{\Sigma}^0-K^-$ molecular picture is gradually
accumulating. However we still urge our experimental colleague
carefully analyze $J/\psi\to p\pi^-\bar{n}$ and $J/\psi\to
p\eta\bar{p}$ channel in further experiments, especially
forthcoming BESIII.

Thus the above analysis shows that the pure molecular state
structure seems to be very difficult to explain $ N_X(1625)$.

We note that there exist two well established states $N^*(1535)$
and $N^*(1650)$ with $J^{P}=1/2^-$ nearby the mass of
${N}_{X}(1625)$. In PDG \cite{PDG}, the branching ratio of
$N^{*}(1650)\to K\Lambda $ is about $3\sim 11\%$. The authors of
Ref. \cite{LIU} indicated that $N^*(1535)$ should have large
$s\bar{s}$ component in its wave function which shows the large
$N^*(1535)K\Lambda$ coupling. $N^*(1535)$ and $N^*(1650)$ can
strongly couple to $K\Lambda$. Thus, before confirming
${N_{X}}(1625)$ to be a new resonance, theorists and
experimentalists of high energy physics need to carry out
cooperation to answer whether $N_X(1625)$ enhancement is related
to $N^*(1535)$ and $N^*(1650)$. Forthcoming BESIII and HIRFL-CSR
will provide the good place to further understand ${N}_X(1625)$
structure.

\begin{widetext}
\begin{center}
\begin{table}[htb]
\begin{tabular}{c|cc}\hline
& $\bar{\Lambda}-K^{-}$ system &$\bar{\Sigma}^{0}-K^{-}$ system\\
\hline $J/\psi\to p \bar N_X(1625)\to p(\pi^{0}\bar{p})$& $1 \times 10^{-8} \sim 3 \times 10^{-8} $& $\sim 1 \times 10^{-3} $ \\
\hline $J/\psi\to p \bar N_X(1625)\to p(\eta\bar{p})$   & $4 \times 10^{-11}\sim 2 \times 10^{-10}$& $\sim 7 \times 10^{-3} $ \\
\hline $J/\psi\to p \bar N_X(1625)\to p(\pi^{-}\bar{n})$& $2 \times 10^{-8} \sim 5 \times 10^{-8} $& $\sim 2 \times 10^{-3} $ \\
\hline
\end{tabular}
\caption{The branching ratios of subordinate decays of $\bar
N_X(1625)$ in two different molecular state pictures.}
\label{numerical}
\end{table}
\end{center}
\end{widetext}

\section*{Acknowledgements}

We thanks Prof. S. Jin, Prof. S.L. Zhu and Dr. Y.R. Liu for useful
discussion. We would like to thank the anonymous referee for many
useful suggestions and comments that have helped me clarify some
points in the original version of the paper. This project was
supported by the National Natural Science Foundation of China
under Grants 10421503, 10625521 and 10705001, Key Grant Project of
Chinese Ministry of Education (No. 305001) and China Postdoctoral
Science foundation (No. 20060400376).

\section*{Appendix}
The further expressions of eqs. (\ref{hadron-loop-1}) and
(\ref{hadron-loop-2}) are
\begin{widetext}
\begin{eqnarray}
\mathcal{M}_{3}^{(\mathcal{A}_3,\mathcal{C}_3)}&=&-g_{3}g_{4}\mathcal{G}
\int^{1}_{0}dx\int^{1-x}_{0}dy
\bigg\{\Big[\frac{(\xi^2-M_{\mathcal{C}_{3}}^{2})y}{16\pi^2
\Delta^2(M_{K},M_{\bar{\Sigma}^{0}},\xi)}\nonumber
-\frac{1}{16\pi^2
\Delta(M_{K},M_{\bar{\Sigma}^{0}},M_{\mathcal{C}_{3}})}\\
\nonumber&&+\frac{1}{16\pi^2
\Delta(M_{K},M_{\bar{\Sigma}^{0}},\xi)}\Big]\bar{v}_{N}\gamma_5
\Big
[\not{p}_{4}\not{p}_{3}[2-2x-(1-x-y)+(1-x-y)x]\nonumber\\&&+\not{p}_{3}\not{p}_{3}[2(1-x-y)-(1-x-y)^2
]+\not{p}_{4}\not{p}_{4}(x-x^2)+\not{p}_{3}\not{p}_{4}(1-x-y)x\nonumber\\&&
+\not{p}_{3}[2M_{\bar{\Sigma}^{0}}-(1-x-y)M_{\bar{\Sigma}^{0}}
]+\not{p}_{4}xM_{\bar{\Sigma}^{0}}\Big]\bar{v}_{{\mathcal{A}_3}}\bigg\}\nonumber\\&&-
g_{3}g_{4}\mathcal{G} \int^{1}_{0}dx\int^{1-x}_{0}dy
\bigg\{\bigg[\frac{2}{(4\pi)^2}\log\Big(\frac{\Delta(M_{K},M_{\bar{\Sigma}^{0}},\xi)}{
\Delta(M_{K},M_{\bar{\Sigma}^{0}},M_{\mathcal{C}_{3}})}\Big)-\frac{(\xi^2-M_{\mathcal{C}_{3}}^{2})y}{8\pi^2
\Delta(M_{K},M_{\bar{\Sigma}^{0}},\xi)}\bigg]\nonumber\\&&
\times\Big[\bar{v}_{N}\gamma_{5}(-\frac{1}{4})\gamma_{\mu}\gamma^{\mu}v_{{\mathcal{A}3}}\Big]\bigg\},
\end{eqnarray}
\begin{eqnarray}
\mathcal{M}_{4}^{(\mathcal{A}_4,\mathcal{C}_4)}&=&-g_{3}'g_{4}'\mathcal{G}
\int^{1}_{0}dx\int^{1-x}_{0}dy
\bigg\{\Big[\frac{(\xi^2-M_{\mathcal{C}_{4}}^{2})y}{16\pi^2
\Delta^2(M_{K},M_{\bar{\Sigma}^{0}},\xi)}\nonumber
-\frac{1}{16\pi^2
\Delta(M_{K},M_{\bar{\Sigma}^{0}},M_{\mathcal{C}_{4}})}\\
\nonumber&&+\frac{1}{16\pi^2
\Delta(M_{K},M_{\bar{\Sigma}^{0}},\xi)}\Big]\bar{v}_{N}\gamma_{5}\Big
[\not{p}_{4}\not{p}_{3}[(1-x-y)-(1-x-y)x]\nonumber\\&&+\not{p}_{3}\not{p}_{3}(1-x-y)^2
+\not{p}_{4}\not{p}_{4}(x^2-x)-\not{p}_{3}\not{p}_{4}(1-x-y)x\nonumber\\&&
+\not{p}_{3}[-M_{\bar{\Sigma}^{0}}(1-x-y)+(1-x-y)M_{\mathcal{C}_4}
]-\not{p}_{4}x(M_{\mathcal{C}_{4}}-M_{\bar{\Sigma}^{0}})+\not{p}_{4}M_{\mathcal{C}_{4}}-M_{\bar{\Sigma}^{0}}M_{\mathcal{C}_{4}}\Big]v_{{\mathcal{A}_4}}\bigg\}\nonumber\\&&-
g_{3}'g_{4}'\mathcal{G} \int^{1}_{0}dx\int^{1-x}_{0}dy
\bigg\{\bigg[\frac{2}{(4\pi)^2}\log\Big(\frac{\Delta(M_{K},M_{\bar{\Sigma}^{0}},\xi)}{
\Delta(M_{K},M_{\bar{\Sigma}^{0}},M_{\mathcal{C}_{4}})}\Big)-\frac{(\xi^2-M_{\mathcal{C}_{4}}^{2})y}{8\pi^2
\Delta(M_{K},M_{\bar{\Sigma}^{0}},\xi)}\bigg]\nonumber\\&&
\times\Big[\bar{v}_{N}\gamma_{5}\frac{1}{4}\gamma_{\mu}\gamma^{\mu}v_{{\mathcal{A}_4}}\Big]\bigg\},
\end{eqnarray}
\end{widetext}
where
\begin{eqnarray*}
\Delta(a,b,c)&=&m_{3}^{2}(1-x-y)^2-2(p_{3}\cdot
p_{4})(1-x-y)x\nonumber\\&&+m_{4}^2
x^2-(m_{3}^{2}-a^{2})(1-x-y)\nonumber\\&&-(m_{4}^{2}-b^{2})x+yc^2,
\end{eqnarray*}
$m_{3}$($m_{4}$) and $p_3$($p_4$) are the masses and  four-
momenta of the final states.

\end{document}